\documentclass[12pt]{article}

\usepackage{rotating}
\usepackage{caption}
\usepackage{amsmath}
\usepackage{amsfonts}
\usepackage[top=0.9in,bottom=0.9in,left=0.8in,right=0.8in]{geometry}
\usepackage{lscape}
\usepackage{float}
\usepackage{booktabs}
\usepackage{url}
\usepackage{color}
 \graphicspath{{figures/}}
\usepackage{graphicx}
\usepackage{bm}
\usepackage[numbers,sort&compress]{natbib}
\newsavebox{\tablebox}

\usepackage{pdfpages}
\usepackage{multirow}
\usepackage{amssymb}
\def\be{\begin{equation}}
\def\ee{\end{equation}}
\def\ben{\begin{equation*}}
\def\een{\end{equation*}}
\def\bea{\begin{eqnarray}}
\def\eea{\end{eqnarray}}
\def\bd{\begin{displaymath}}
\def\ed{\end{displaymath}}
\def\bda{\begin{eqnarray*}}
\def\eda{\end{eqnarray*}}

\def\ha1{\hat \beta_1}

\def\bsc{\begin{scriptsize}}
\def\esc{\end{scriptsize}}

\begin{document}
\begin{Large}
\title{\bf Data-driven research on chemical features of Jingdezhen and Longquan celadon by energy dispersive X-ray fluorescence}
\author{Ziyang He\textsuperscript{a,$\dagger$,$\ast$}, Maolin Zhang\textsuperscript{b,$\dagger$}, Haozhe Zhang\textsuperscript{c}\\
\\
\small { \textsuperscript{a}Department of Industrial and Manufacturing System Engineering,}\\
\small{Iowa State University, Ames, IA 50011 USA.}\\
\small{  \textsuperscript{b}Jingdezhen Ceramic Institute, Jingdezhen 333001, China.}\\
\small{  \textsuperscript{c}Department of Statistics, Iowa State University, Ames, IA 50011 USA}
}
\date{}
\maketitle

\end{Large}
\noindent{\textsuperscript{$\dagger$} Z.H. and M.Z. are joint first authors.}\\

\noindent{\textsuperscript{$\ast$}Corresponding Author:\\ 
Ziyang He\\
       Email: heziyang@iastate.edu\\
       Address: \\
       Department of Industrial and Manufacturing System Engineering, \\
       Iowa State University, Ames, Iowa 50014, USA.\\}

\vspace{4.5\baselineskip}
\newpage
\noindent{\bf Abstract}
The energy dispersive X-ray fluorescence (EDXRF) is used to determine the
chemical composition of celadon body and glaze in Longquan kiln (at
Dayao County) and Jingdezhen kiln. Forty typical shards in four cultural eras were selected to investigate the raw materials
and firing technology. Random forests, a relatively new statistical technique, has been adopted to identify chemical elements that are strongest explanatory variables to classify samples into different cultural eras and kilns. 
The results indicated that the contents of Na$_{2}$O,
Fe$_{2}$O$_{3}$, TiO$_{2 }$, SiO$_{2 }$ and CaO vary in celadon bodies from Longquan and
Jingdezhen, which implies that local clay was used to manufacture celadon bodies in Jingdezhen kiln.
By comparing the chemical composition in glaze, we find that the chemical elements and firing
technology of Jingdezhen kiln are very similar to those in Longquan kiln, especially for Ming dynasty.
This study reveals the inheritance between Jingdezhen kiln and Longquan kiln, and explains the differences between those two kilns.


\bigskip

\section{Introduction}
Celadon is one of the earliest porcelains in China as well as in the world.
With its long history and rich varieties, celadon has attracted a lot of attentions from researchers in archaeology, physics, material science, etc. Longquan kiln, one of the most
representative kilns in celadon production, enjoys an important
status in Chinese ceramic history \cite{1}. The government in Ming Dynasty invested
nationwide labors, resources and capitals in the establishment of imperial kiln in
Jingdezhen from Yongle era. Jingdezhen has also became the craftsmen aggregation center and the porcelain manufacturing center in Maritime Silk Road since Yongle era of Ming Dynasty \cite{2}. Jingdezhen kiln were able to make color glaze and whiteware porcelains from Yuan Dynasty and began to make celadon from Tang Dynasty. The demand from commercial activities, foreign trade and imperial reward far stripped the supply of Jingdezhen porcelain \cite{3}. Due to the good quality and unique glaze feature (jade texture) of Longquan celadon, Jingdezhen kiln was trying to imitate Longquan celadon from style, color and pattern. The producing and firing technology of imitated Longquan celadon in Jingdezhen had reached a relative high level in Yongle era of Ming Dynasty (1403 - 1424), however craftsmen did not master the completely mature firing technology until Yongzheng era of Qing Dynasty (1723 - 1735) \cite{4}. At present, Longquan celadon has been in depth studied by current archaeologist while the imitated Longquan celadon
in Jingdezhen still remains little investigated.

It has been well known that the contents
of macro and micro elements in the body and glaze of porcelain are dependent on
the raw material and firing technology \cite{5,6}. Li et al. \cite{7} used
EDXRF to confirm that the raw materials of imperial Longquan porcelain
have no obvious changes between Hongwu and Yongle era (1368 - 1398), and claimed the
firing technology has not declined in the Early Ming Dynasty. Zhu et al. (2011)
\cite{8} shows micro elements, e.g. Cr, Sr, Zr in body and Sr,
Rb in glaze could be employed as relevant markers for the non-destructive
discrimination of the provenance of Xicun and Yaozhou kiln. Wu et al. (2015) \cite{9}
 is a research on the early celadon of Jingdezhen and its
initial development. It comparatively analyzed the EDXRF data from Nan and Lantin kiln sites in Jingdezhen and explained
the difference in chemical composition and processing characteristics.

In this study, the samples of body and glaze of Longquan celadon in Northern Song, Southern Song, 
Yuan and Ming Dynasties as well as imitated Longquan celadon in Jingdezhen
in Ming Dynasty were examined by the energy dispersive X-ray fluorescence (EDXRF) microprobe. EDXRF is an efficient
instrument that combines the merits of X-ray fluorescence spectrometry with the ability to analyze a
micro-area. The large chamber size allows for the non-destructive analysis
of macro and trace elements on micro-areas of the samples. After obtaining the data, Random forests, a well-established machine learning algorithm, was adopted in the analysis, rather than multivariate statistical methods. If many elements were measured in the experiments, it's very inefficient to find the difference 
by standard multivariate statistical methods (Li et al. 2010 \cite{10}). However, Random forests can return the values that measure the variable importance of explanatory variables in the training data, so only the top-ranked elements need to be investigated. The results of the analysis reveal the inheritance relationship of the celadon from Longquan and Jingdezhen kiln.

\section{Experiment}
\begin{figure}[H]
  \centering
  \includegraphics[width=0.55\textwidth]{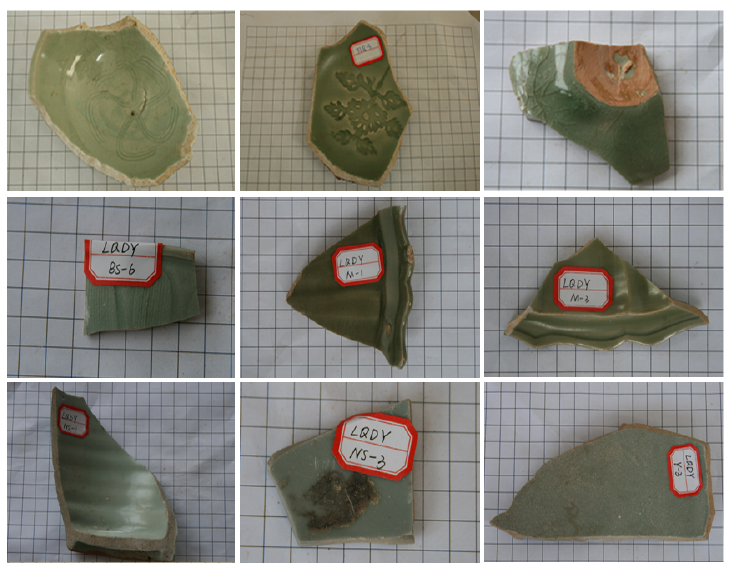}
  \caption{Photographs of celadon samples from the Longquan
kiln and Jingdezhen kiln}\label{fig:2}
\end{figure}

In this experiment, 40 samples of Longquan celadon in Northern Song, Southern Song, Yuan and Ming
Dynasties as well as imitated Longquan celadon in Jingdezhen civilian kilns
in Ming Dynasty are collected. They include 7 samples of Longquan celadon in
Northern Song Dynasty (LQ-BS-1$\sim $LQ-BS-7), 8 samples of Longquan celadon
in Southern Song Dynasty(LQ-NS-1$\sim $LQ-NS-8), 9 samples of Longquan
celadon in Yuan Dynasty(LQ-Y-1$\sim $LQ-Y-9), 3 samples of Longquan celadon
in Ming Dynasty(LQ-M-1$\sim $LQ-M-3) and 13 samples of imitated Longquan
celadon in Jingdezhen in Ming Dynasty(FLQ-M-1$\sim $FLQ-M-13). The
photographs of samples are shown in Figure \ref{fig:2}.

The samples were cut and cleaned in an ultrasonic bath and then dried before
testing. Energy Dispersive X-ray Fluorescence (EDXRF) spectrometer
(EAGLE-III) was used to detected body and glaze compositions. The analysis
was executed at 50 kV and 200 mA voltage-current of the X-ray tube, with a
vacuum optical route and dead time was around 25{\%}. The detector is a
liquid-nitrogen-cooled Si (Li) crystal with Rh window, and the beam spot was
300 $\mu $m. The software employed for spectrum retraction and analysis was
the program VISION32, associating with the instrument. Quantitative analysis
was operated by calibration curve method. The calibration samples were from a
set of reference samples (13 pieces) with known chemical compositions,
provided by the Shanghai Institute of Ceramics of the Chinese Academy of
Science (SICCAS). The analytical results were in Table \ref{tab:1} and \ref{tab:4}.

The firing temperatures of the typical samples were estimated from the
inflection point of the thermal expansion curves (Table \ref{tab:2}) and 
measured by DIL 402C Thermal Dilatometer of the German NETZSCH Instrument
Company. Water absorptions were tested through boiling method.

\section{Results and discussion }

\subsection{Random forests and variable importance}
Random forests,  an increasingly popular
nonparametric methodology, is an extension of classification and 
regression trees (CART) method \cite{11}. It grows many classification trees or
regression trees and thus has the name "forests". Every tree is built using a
deterministic algorithm, and the trees are different in two aspects:
First, at each node, a best split is chosen from a random subset of the
predictors rather than all of them; Second, every tree is built using a
bootstrap sample of the observations. The remaining sample, the so-called "out-of-bag" (OOB) sample,
which contains approximately one-third of the observations, are then used to estimate the
prediction accuracy. A key feature of random forest is its ability to measure variable importance \cite{12}. Variable importance reflects the degree
of association between a given explanatory variable and the response variable. More details about random forests and its wide applications can be found in \cite{13, 14}.

In this research, the contents of chemical elements are the explanatory variables and the categories of eras and kilns are the response variable. By using \textit{randomForest} Package in R language \cite{15}, we performed random forests method to do classification of celadon samples.  Mean descrease in accuracy (MDA) and mean decrease in Gini index (MDG), two values that random forests method returns after constructing classification trees, can be used to identify chemical elements that differ most significantly among goups of samples in different categories of ears and kilns. The returned values of MDA and MDG are shown in Table \ref{vi_body} and \ref{vi_glaze}. Random forests method is more efficient in selecting important variables than standard multivariate statistical methods, such as principal component analysis and linear regression, especially when the number of measured chemical elements is large \cite{10}.

\subsection{Analysis of the chemical composition of the celadon body in
Longquan and Jingdezhen}

The data in Table \ref{tab:1} show obvious differences in compositions of bodies between
Longquan and Jingdezhen kilns. It can be seen that the Al$_{2}$O$_{3
}$ contents of celadon body in Jingdezhen (19.62{\%}) are lower than those in
Longquan (22.90{\%}), whereas the SiO$_{2}$ contents of celadon body in
Jingdezhen (72.40{\%}) are higher than those of the Longquan (67.93{\%}).
The average contents of (RO + R$_{2}$O) are 5.70{\%} in Jingdezhen and
6.27{\%} in Longquan. The flux contents in two kilns don't display any significant
differences. The only difference is that Jingdezhen kiln is in high silicon
flux domain, but Longquan kiln is in low silicon flux domain. Meanwhile,
together with results from sintering temperature test (see Table \ref{tab:2}), it can
been shown that the average sintering temperature of celadon is (1139.75 $\pm
$ 20) $^{\circ}$C in Jingdezhen kiln and mainly between 1157 $^{\circ}$C and 1230 $^{\circ}$C
in Longquan kiln. In order to meet the demand of fully sintered body,
average sintering temperature of celadon in Jingdezhen is slightly lower
than that in Longquan.

In our research, EDXRF data of the bodies in different cultural eras are investigated. We did leave-one-out cross-validation \cite{16} of chemical composition data by using Random forests. The first main conclusion is that there is no obvious difference among different cultural eras in Longquan kiln, since the mis-classification ratio of cross-validation is 52.2{\%}. Another
important conclusion is that low mis-classification ratio (2.6{\%}) shows clear
difference between samples from two kilns. Table \ref{vi_body} shows variable importance of observed data. In macro-element domain, Fe$_{2}$O$_{3}$ (MDG = 3.7851) and CaO (MDG = 2.7215)
are two most important elements to classify origins of products in Longquan and Jingdezhen. In the micro-element domain, ZrO$_{2}$ (MDG = 4.0653) and Y$_{2}$O$_{3}$ (MDG = 2.3640) are other two important elements. 

\begin{figure}[H]
  \centering
  \includegraphics[width=0.65\textwidth]{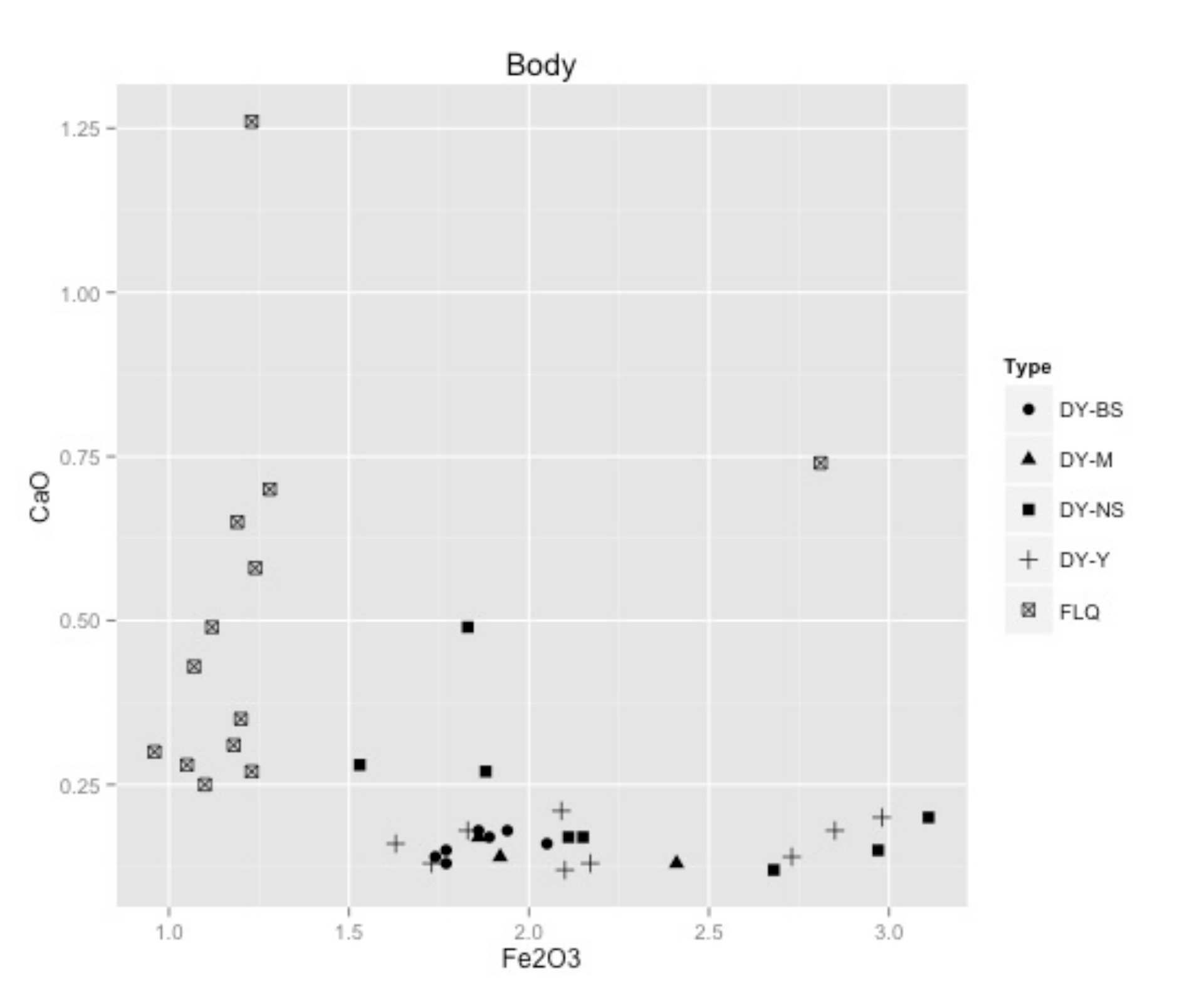}
  \caption{The scatterplot of macro-elements in the bodies
in different cultural eras and kilns.}\label{fig:4}
\end{figure}

As displayed in Figure \ref{fig:4}, the celadon bodies from Jingdezhen contain more CaO
than Longquan celadon and have a relatively large fluctuating range. The bodies from Longquan own higher contents of Fe$_{2}$O$_{3}$ than those bodies from Jingdezhen and also have a wide fluctuating range. Bodies sampled from Jingdezhen were dispersed in lower left domain, while bodies sampled from Longquan were in lower right domain. The contents of Fe$_{2}$O$_{3}$ in Longquan samples in different cultural eras can show the continuous inheritance relationship of body formula evolution. It has been confirmed by the data from archeological research that Southern Song Dynasty kiln site is in the vicinity of Ming Dynasty kiln site \cite{17}. This has also indicated that the albite contents are high in raw materials of Jingdezhen kiln whereas potassium feldspars are the main raw materials of Longquan kiln, which can directly be seen from Table \ref{tab:3} \cite{18} \cite{19}.

The iron and titanium contents are usually higher in samples from Longquan kiln than those from
Jingdezhen kiln, as can be seen in Table \ref{tab:3}, raw materials in Jingdezhen are featured by lower iron and titanium contents. The average Fe$_{2}$O$_{3}$ content in
Mingsha kaolin soil and Xingzi kaolin soil is 1.335{\%} and the average
TiO$_{2}$ content is 0.025{\%}, which are much lower than those of Longquan
kiln soil (Taking Zijin soil in the Dayao County Gaojitou village as an
example). The average Fe$_{2}$O$_{3}$ content is 3.11{\%} and average
TiO$_{2}$ content is 0.45{\%}, respectively. From the appearance of the samples,
most celadon bodies from Longquan are gray while bodies from Jingdezhen are
mainly white.

In conclusion, the main reasons for the differences mentioned above can be listed as followed.
Firstly, the body element is influenced by local raw materials and addition
of Zijin soil. Compared with Jingdezhen kiln, Longquan kiln displays the
features of high aluminum and iron with low silicon. This is mainly due to the fact that Zijin soil has high contents of iron and aluminum was added during bodies
manufacturing. Secondly, the selection of raw materials for making celadon
bodies is limited by local resources and has to tailor to local conditions.
From excavated archeological data \cite{17}, the raw materials for making
porcelain was obtained in the vicinity of the two kiln sites, thus raw
materials in distinct kilns remained different. With consideration for cost,
the craftsmen in Jingdezhen were unable to acquire the celadon body raw
materials in great demand from distant areas. Consequently, at that time
they employed the local raw materials to imitate Longquan celadon, which
resulted in the differences in chemical compositions in two kilns'
production.

\subsection{Analysis of the chemical composition of the celadon glaze in
Longquan and Jingdezhen}
Longquan celadon glaze can be roughly divided into two categories according
to the features. One category is the transparent glaze with high gloss,
which could be represented by the glaze in Northern Song Dynasty; the other
category is the jade texture glaze, which could be represented by the glaze
in Southern Song Dynasty \cite{20}. Glaze from Jingdezhen kiln shares similar
features. From Table \ref{tab:4}, the CaO contents of Longquan celadon in Northern
Song Dynasty (10.75{\%}) were significantly higher than those of the
Jingdezhen in Ming Dynasty (6.20{\%}) as well as those of Longquan kiln in
other cultural eras (7.47{\%}), while R$_{2}$O contents are significantly lower and only
reach 4.48{\%}. As Seen from the samples appearances, the glaze of Longquan
celadon in Northern Song Dynasty is glossier and more transparent. Carved
patterns can be clearly seen and these samples do have smoke absorption
phenomenon.

As aforementioned, random forests has also been used to analyze EDXRF data of glaze samples. 
There is still no obvious difference between different
cultural eras in glaze samples from Longquan kiln, since the mis-classification ratio of cross-validation is 
53.2{\%}.  
Table \ref{vi_glaze} shows variable importance of chemical elements in glaze. In marco-element region, Na$_{2}$O (2.1794) and CaO (1.9602) are the two most important macro-elements; SrO (4.3968) and ZrO$_{2}$ (3.3247) are two other important micro-elements variables. Consitent with that Li et al. (2010) \cite{10} claiming Sr isotopic and trace chemical features of whiteware shards can be linear array rule to distinguish Nanwan and three other sites in the Erlitou culture period.

\begin{figure}[H]
  \centering
  \includegraphics[width=0.6\textwidth]{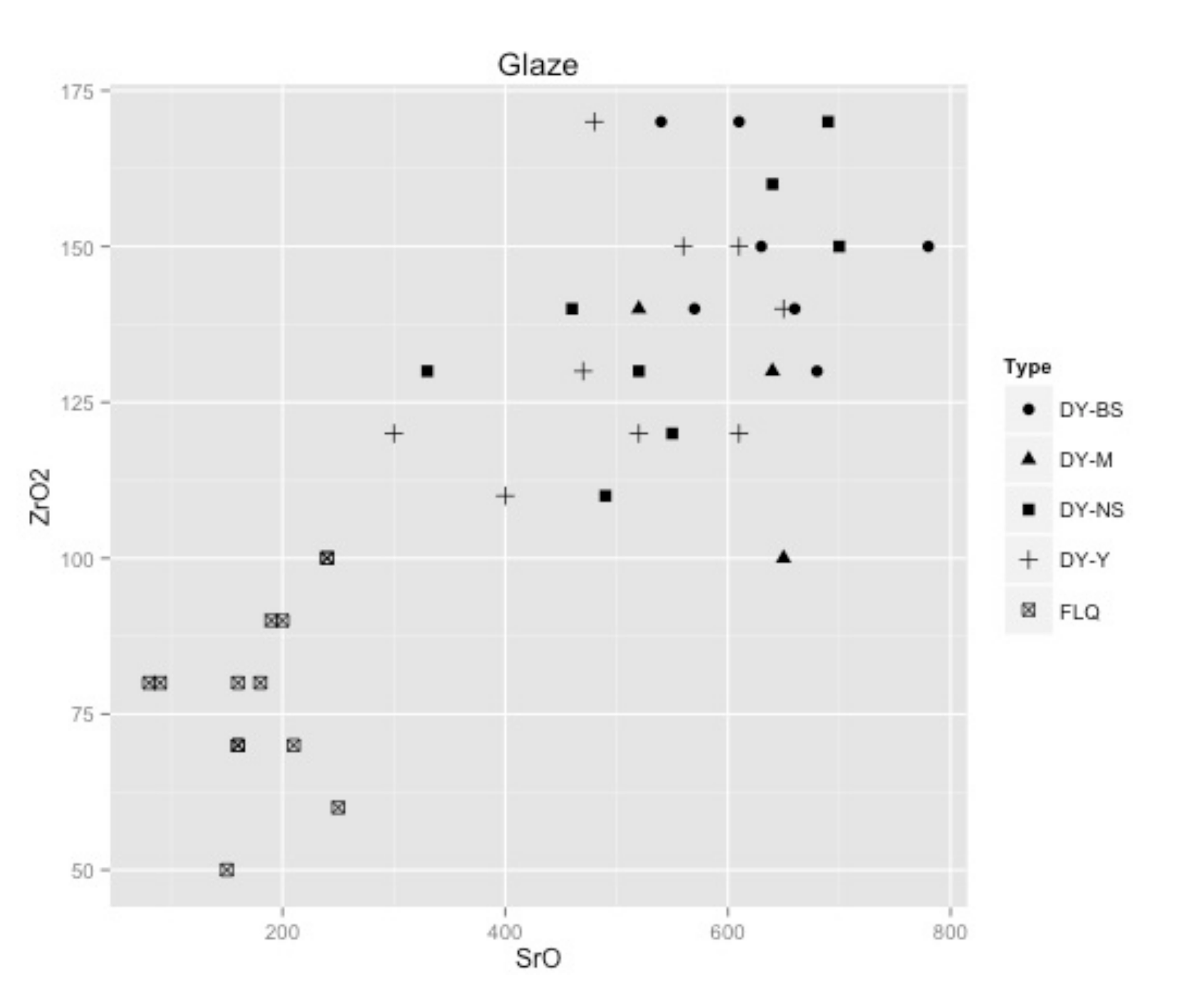}
  \caption{The scatterplot of micro-elements of the glaze in
different cultural eras and kilns}\label{fig:7}
\end{figure}

\begin{figure}[H]
  \centering
  \includegraphics[width=0.65\textwidth]{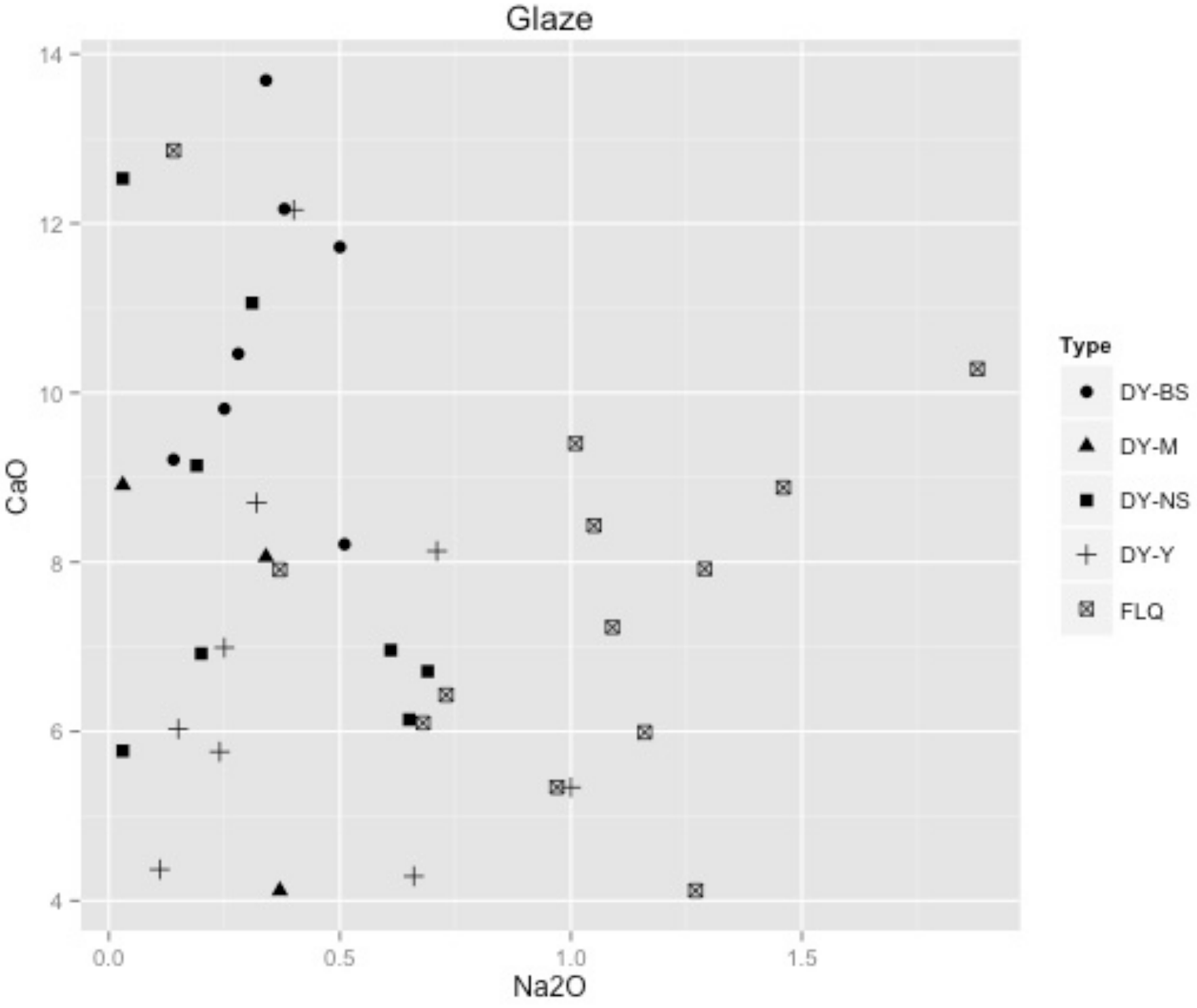}
  \caption{The scatterplot of macro-elements of the glaze in
different cultural eras and kilns}\label{fig:6}
\end{figure}

From the scatterplot of SrO and ZrO$_{2}$ (Figure \ref{fig:6}), it can be
seen that the points of samples from Longquan kiln are in the right upper part and the points of samples from Jingdezhen kiln are in left lower part. Compared with Figure \ref{fig:7}, the macro-element CaO and Na$_{2}$O domains of Longquan kiln have some overlapping parts with Jingdezhen kiln, especially in Southern Song and
Yuan Dynasty. This indicated the formulas the craftsmen in Jingdezhen were using to imitate, and micro-element level could tell the difference of these formulas, which is reasonable to believe
due to different origins and technical modification.

From Southern Song Dynasty to Ming Dynasty, R$_{2}$O contents of Longquan
celadon show obvious increase while CaO contents display gradual reduction
(Table \ref{tab:4}). R$_{2}$O contents of Longquan celadon reached a peak (5.98{\%}) in
Ming Dynasty with the CaO contents of 6.20{\%}. It indicated that the chemical composition of
Longquan celadon glaze had successfully transformed from calcium glaze to
calcium-alkali glaze \cite{21}.

Furthernore, compared with the high similarity (cultural eras continuity) among
the glaze samples from Longquan kiln, the imitated celadon glaze from Jingdezhen can be divided into three categories \cite{17}. 
The proper growth in N$_{2}$O contents can
lower the melting temperature of glaze, broaden the melting temperature
variation and increase the high temperature viscosity to retain more bubbles
and un-melted quartz which could improve the jade texture of celadon \cite{6}. At the
same time, the firing temperature of celadon with better jade texture is
usually at the lower bound of positive combustion. Once the firing
temperature enters the over firing temperature variation, the glaze will
become transparent and glossy and lose its jade texture \cite{20}. As can be seen
in Table \ref{tab:3}, the firing temperature of Jingdezhen celadon is around 1140\r{
}C, this is much lower than that of ordinary ceramics (1250\r{ }C) in
Jingdezhen and a bit lower than that of Longquan celadon (1174\r{ }C). The assumption is the craftsmen made some adjustments on purpose to
develop a better jade texture for the glaze due to the uniqueness of
celadon. The average contents of alkali metal oxides in
imitated celadon in Jingdezhen are clearly higher than those of Longquan
celadon. Some samples (FLQ-12) with high R$_{2}$O have irregular cracks in the
glaze layers, mainly due to the relatively high Na$_{2}$O contents in the glaze \cite{22}.

\begin{figure}[H]
  \centering
  \includegraphics[width=0.65\textwidth]{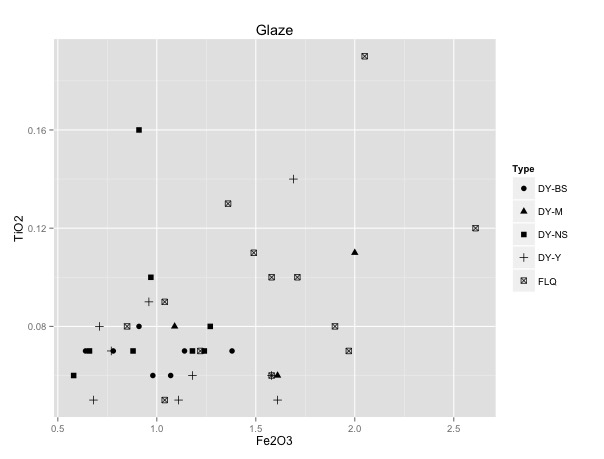}
  \caption{The scatterplot of Fe$_2$O$_3$ and TiO$_2$ in different cultural eras and kilns}\label{fig:8}
\end{figure}

As shown in Figure \ref{fig:8}, the contents of Fe$_{2}$O$_{3 }$ and TiO$_{2 }$
in celadon samples from Jingdezhen are very similar but much higher than
those of Longquan celadon in Northern Song and Southern Song Dynasty. This reveals that the craftsmen at that time knew the formula of 
Longquan celadon well and also found alike raw materials \cite{23}, for instance,
the Zijin soil with high contents of iron. In the meantime, the Na$_{2}$O
contents (1.01{\%}) in imitated Longquan celadon in Jingdezhen in Ming
Dynasty are different from those in Longquan celadon in Ming Dynasty (Figure
\ref{fig:8}), which further explains that the Jingdezhen adopted a kind of local raw
materials that are not completely same with that in Longquan.

\section{Conclusion }
By analyzing data of body and glaze chemical
compositions, we obtained the following conclusions. There are some differences in the body chemical compositions between the
imitated Longquan celadon in Jingdezhen and Longquan celadon. The body of
imitated Longquan celadon in Jingdezhen has comparatively high contents of
silicon and low contents of iron and titanium, which is similar to the
feature of raw material in Jingdezhen location. It shows that Jingdezhen
craftsmen acquired the celadon bodies' raw materials in great demand from
the local area during the imitated Longquan celadon producing.

Imitated Longquan celadon in Jingdezhen directly or indirectly learned
producing technology from Longquan celadon. Jingdezhen celadon is similar to
Longquan celadon both in glaze chemical composition and firing technology.
Jingdezhen craftsmen not only sought for the similar glaze raw materials in
local area but also employed the firing temperature of around 1140\r{ }C to
produce the imitated glazes. This firing temperature is close to that of
Longquan celadon (1174\r{ }C) but obviously much lower than that of ordinary
ceramics (1250\r{ }C) in Jingdezhen.

The chemical compositions of imitated Longquan celadon glaze in Jingdezhen
can be divided into three categories. One has much higher contents
of alkali metal oxide in some samples than Longquan celadon glaze. They have
more cracks in the glaze layers. The other two categories are respectively
consistent with the features of Longquan celadon glaze in Northern Song,
Southern Song and Ming Dynasties.

\section*{Acknowledgements}
M. Zhang's research was partially supported by the National Science Foundation of China
(51362017 and 11205073) and Social Science Funds of Jiangxi Province
(2010GZC0088). 

\newpage
\noindent
\section*{\bf Figures and tables}
\begin{figure}[H]
  \centering
  \includegraphics[width=0.95\textwidth]{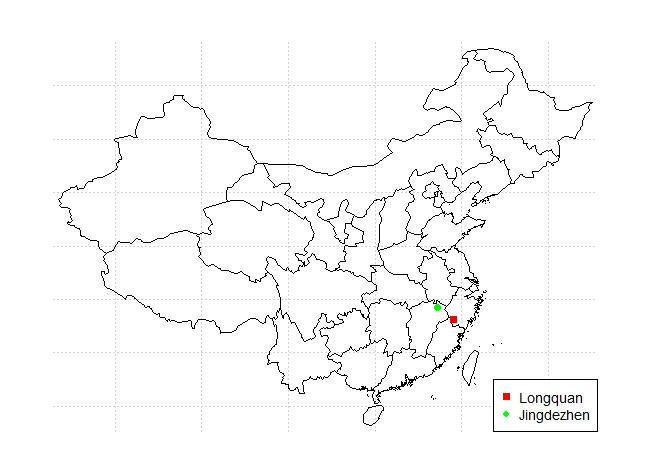}
  \caption{The locations of Longquan kiln (Dayao County) and Jingdezhen kiln}\label{fig:1}
\end{figure}

  \newpage
  
\begin{table}[H]
  \centering
    \small
     \caption{Chemical composition of the bodies from Longquan and Jingdezhen kilns}
  \tabcolsep = 2pt
\begin{tabular}{llllllllllllll}
\toprule
Group&
Na$_2$O (\%)&
MgO (\%)&
Al$_2$O$_{3}$ (\%)&
SiO$_2$ (\%)&
K$_2$O (\%)&
CaO (\%)&
TiO$_2$ (\%)&
Fe$_2$O$_3$ (\%)&
\\

\midrule
FLQ-M&
0.52(0.21)&
0.39(0.15)&
19.62(1.52)&
72.40(2.10)&
4.20(0.53)&
0.51(0.28)&
0.08(0.05)&
1.28(0.47)\\
LQ-BS&
0.30(0.22)&
0.33(0.11)&
21.71(2.15)&
68.99(2.82)&
5.51(0.51)&
0.16(0.02)&
0.14(0.03)&
1.86(0.11)\\
LQ-NS&
0.35(0.19)&
0.36(0.12)&
22.62(2.30)&
67.85(2.72)&
5.16(0.76)&
0.23(0.12)&
0.15(0.08)&
2.28(0.57)\\
LQ-Y&
0.37(0.21)&
0.36(0.11)&
23.55(1.74)&
67.06(2.26)&
5.13(1.07)&
0.16(0.03)&
0.15(0.08)&
2.23(0.50) \\
LQ-M&
0.30(0.12)&
0.19(0.02)&
23.23(1.13)&
67.44(1.83)&
5.54(0.48)&
0.15(0.03)&
0.09(0.02)&
2.06(0.30)\\
\bottomrule
\end{tabular}
\label{tab:1}
\end{table}

\vspace{4\baselineskip}

\begin{table}[H]
  \centering
  \caption{Chemical composition of the glaze from Longquan and Jingdezhen kilns}
  \small
  \tabcolsep = 3pt
\begin{tabular}{llllllllllllll}
\toprule
Group&
Na$_2$O (\%)&
MgO (\%)&
Al$_2$O$_{3}$ (\%)&
SiO$_2$ (\%)&
K$_2$O (\%)&
CaO (\%)&
TiO$_2$ (\%)&
Fe$_2$O$_3$ (\%)&
\\
\midrule
FLQ-M&
1.01(0.46)&
0.41(0.25)&
12.87(0.67)&
70.21(2.16)&
5.08(0.95)&
7.76(2.31)&
0.09(0.04)&
1.57(0.49)\\

LQ-BS&
0.34(0.14)&
0.65(0.25)&
13.42(1.19)&
68.64(1.81)&
4.13(0.57)&
10.75(1.89)&
0.07(0.01)&
0.99(0.24)\\

LQ-NS&
0.34(0.27)&
0.58(0.22)&
13.56(0.65)&
70.43(1.84)&
4.89(0.91)&
8.15(2.49)&
0.08(0.03)&
0.96(0.26)\\

LQ-Y&
0.43(0.19)&
0.47(0.05)&
12.84(1.13)&
71.81(2.43)&
5.38(1.12)&
6.86(3.22)&
0.07(0.04)&
1.14(0.43) \\

LQ-M&
0.31(0.12)&
0.55(0.27)&
12.90(6.78)&
72.13(40.15)&
5.67(2.54)&
6.20(1.94)&
0.07(0.02)&
1.17(0.42) \\
\bottomrule
\end{tabular}
\label{tab:4}
\end{table}

\begin{table}[H]
  \centering
  \caption{Sintering temperature of typical samples of celadon from
Jingdezhen and Longquan kilns}
    \begin{tabular}{llc}
    \toprule
    Number & Kiln Location & Sintering Temperature ($^{\circ}C$) \\\midrule
    LQ-BS-3 & Longquan (Dayao)        & 1182 $\pm$ 20 \\
    LQ-NS-1 & Longquan & 1180 $\pm$ 20 \\
    LQ-NS-1 & Longquan (Dayao)        & 1206 $\pm$ 20 \\
    LQD-01 & Longquan (Dayao)        & 1181 $\pm$ 20 \\
    LQD-04 & Longquan (Dayao)        & 1171 $\pm$ 20 \\
    LQ-Y-1 & Longquan (Dayao)        & 1073 $\pm$ 20 \\
    LQD-07 & Longquan (Dayao)        & 1187 $\pm$ 20 \\
    LQ-M-2 & Longquan (Dayao)        & 1157 $\pm$ 20 \\
    ML-1  & Longquan & 1230 $\pm$ 20 \\
    FLQ-3 & Jingdezhen & 1159 $\pm$ 20 \\
    FLQ-5 & Jingdezhen & 1167 $\pm$ 20 \\
    FLQ-8 & Jingdezhen & 1132 $\pm$ 20 \\
    FLQ-12 & Jingdezhen & 1101 $\pm$ 20 \\
    \bottomrule
    \end{tabular}%
  \label{tab:2}%
\end{table}%

\begin{table}[H]
\centering
  \small
  \caption{Variable importance for elements in body }
\begin{tabular}{ccc}
  \hline
Element & MDA & MDG \\ 
  \hline
ZrO$_{2}$ & 0.0704 & 4.0653 \\ 
  Fe$_{2}$O$_{3}$ & 0.0515 & 3.7851 \\ 
  CaO & 0.0398 & 2.7215 \\ 
  Y$_{2}$O$_{3}$ & 0.0386 & 2.3640 \\ 
  MgO & 0.0012 & 1.9451 \\ 
  Rb$_{2}$O & 0.0193 & 1.9348 \\ 
  SiO$_{2}$ & 0.0067 & 1.7819 \\ 
  K$_{2}$O & 0.0080 & 1.5509 \\ 
  Al$_{2}$O$_{3}$ & 0.0050 & 1.5311 \\ 
  P$_{2}$O$_{5}$ & 0.0022 & 1.4201 \\ 
  Na$_{2}$O & -0.0016 & 1.2934 \\ 
  MnO & -0.0008 & 1.2618 \\ 
  TiO$_{2}$ & -0.0004 & 1.2286 \\ 
  PbO$_{2}$ & -0.0021 & 1.1083 \\ 
  ZnO & 0.0013 & 0.8112 \\ 
  CuO & -0.0042 & 0.7016 \\ 
  SrO & -0.0015 & 0.4448 \\ 
   \hline
\end{tabular}  \label{vi_body}
\end{table}

\begin{table}[H]
\centering
\small
\caption{Variable importance for elements in glaze}
\begin{tabular}{ccc}
  \hline
 Element& MDA & MDG \\ 
  \hline
SrO & 0.0781 & 4.3968 \\ 
  ZrO$_{2}$ & 0.0644 & 3.3247 \\ 
  Rb$_{2}$O & 0.0235 & 2.2258 \\ 
  Na$_{2}$O & 0.0133 & 2.1794 \\ 
  CaO & 0.0047 & 1.9602 \\ 
  MnO & 0.0110 & 1.9423 \\ 
  P$_{2}$O$_{5}$ & 0.0084 & 1.7970 \\ 
  K$_{2}$O & 0.0033 & 1.6330 \\ 
  Fe$_{2}$O$_{3}$ & 0.0008 & 1.4497 \\ 
  SiO$_{2}$ & -0.0002 & 1.4034 \\ 
  Al$_{2}$O$_{3}$ & -0.0023 & 1.3054 \\ 
  MgO & -0.0017 & 1.2400 \\ 
  Y$_{2}$O$_{3}$ & 0.0092 & 1.1379 \\ 
  ZnO & 0.0008 & 1.1042 \\ 
  TiO$_{2}$ & 0.0036 & 1.0291 \\ 
  CuO & -0.0014 & 0.9623 \\ 
  PbO$_{2}$ & -0.0012 & 0.7448 \\ 
   \hline
\end{tabular} \label{vi_glaze}%
\end{table}

\begin{table}[H]
  \centering
  \caption{Chemical composition of refined mud from Longquan and Jingdezhen}
  \footnotesize
  \tabcolsep = 5.5pt
    \rotatebox{90}{
    \begin{tabular}{cllllllllllll}
    \toprule
    Location & Name  & SiO$_2$  & Al$_2$O$_3$ & CaO   & MgO   & K$_2$O   & Na$_2$O  & Fe$_2$O$_3$ & TiO$_2$  & MnO   & Loss  & Total \\
    \midrule
          & Dayao soil & 71.66 & 17.96 & 0.01  & 0.22  & 2.13  & 0.16  & 1.63  & ---     & 0.02 & 6.06 & 99.85 \\
    Longquan & Gaojitou Zijin soil & 66.93 & 18.01 & 1.23  & 0.51  & 5.23 & 0.45  & 3.11  & 0.45  & 0.08 & 4.47 & 100.47 \\
          & Huangliankeng Zijin soil & 45.92 & 24.77 & 0.46  & 0.86  & 1.53  & 0.53 & 13.85 & 2.00 &--- & 10.38 & 99.30 \\
          & Sanbaopeng chinastone  & 73.70 & 15.34 & 0. 70 & 0.16 & 4.13  & 3.79 & 0.70 & --- & 0.04 & 1.13  & 99.69 \\\midrule
    \multirow{2}[0]{*}{Jingdezhen } & Nangang chinastone & 76.12 & 14.97 & 1.45  & ---     & 2.77 & 0.42  & 0.76 & ---     & 0.06 & 3.71  & 100.26 \\
          & Xingzi kaolin soil & 51.89 & 31.7  & 0.91  & Trace & 2.05 & 2.05 & 1.54  & ---     & 0.82 & 11.01 & 99.97 \\
          & Mingsha kaolin soil & 49.65 & 33.82 & 0.33 & 0.23  & 2.70 & 1.03  & 1.13  & 0.05 & 0.33 & 10.84 & 100.11 \\
    \bottomrule
    \end{tabular}%
    }
  \label{tab:3}%
\end{table}%

\end{document}